\documentclass[reprint, amsmath, amssymb, aps, prd]{revtex4-1}

\usepackage[english]{babel}
\usepackage[utf8]{inputenc}
\usepackage[normalem]{ulem} 
\usepackage[colorlinks=true, citecolor=blue, filecolor=magenta, unicode]{hyperref}
\usepackage[capitalise]{cleveref}

\usepackage{dcolumn}
\usepackage{graphicx}
\usepackage{xcolor}
\usepackage{bm}
\usepackage{amsmath}
\usepackage{amssymb}
\usepackage{arydshln}
\usepackage{mathtools}
\usepackage{soul}
\usepackage[caption=false]{subfig}

\graphicspath{ {./fig/}{./pictures/} }
\newcommand{\bref}[1]{(\ref{#1})}
\DeclareMathOperator{\arcosh}{arcosh}

\begin{document}

\title{New method of accounting for interference contributions \\ within a multipheriperal model}

\author{O.~Potiienko}
\email{frumle@ukr.net}       
\author{K.~Merkotan}
\author{N.~Chudak}
\author{D.~Ptashynskiy}
\author{T.~Zelentsova}
\author{T. Yushkevich}
\author{I.~Sharph}
\author{V.~Rusov}
\affiliation{ Odessa National Polytechnic University \\
              Schevchenko av. 1, Odessa, 65044, Ukraine }

\date{\today}

\begin{abstract}
We consider an inelastic scattering of protons within the simplest real scalar model $\phi^3$ (phi-cubed). Although this model is being studied for a very long time, the problem of accounting for the interference contributions for all the possible particle multiplicities observed in experiment is not solved yet. We propose the method which is based on grouping of the interference contributions into sets in such a way that the sum of all interference contributions of each particular set can be calculated with Laplace's method. This approach allowed us to calculate all the interference contributions to the cross-sections for multiplicities up to  $ n\sim 50 $ at the energy $ \sqrt{s} \sim 50$ GeV. The obtained models of the energy dependence of total $pp$ scattering cross-section and the inclusive rapidity distribution are in qualitative agreement with the experiment.

We also consider the well known effect of the energy dependence of the shape of inclusive rapidity distribution, and propose an explanation of this dependence and consider it exactly as the interference effect. 
\end{abstract}

\maketitle

\section{Introduction}
\label{sec:intro}
When calculating the cross-sections of an inelastic protons scattering within the multiperipheral model \cite{Amati1962,Kuraev:1977fs} it is common to make an assumption that the multi-Regge region \cite{BAKER19761, Lipatov:2008,MultiRggeBFKL,KRF, Duhr2019} makes a main contribution to the multidimensional integral. According to this assumption, the produced particles are strongly ordered in rapidity. As a result, the Reggeization models \cite{Amati1962,Kuraev:1977fs,FADIN197550,KRF} use the approximations so that the energy-momentum varies widely within the integration domain (see the last section in \cite{Sharph:2011wm}). At the same time it is possible to specify without any assumptions the integration domain which makes a main contribution to the multidimensional integrals for the scattering cross-section. It has been shown (see \cite{Sharf:2006it, Sharf:2007cf, Sharf:2009yy, Sharf:2011zz, Sharph:2011wm, Sharf:2013ona}) that the squared modulus of the multipheriperal diagram contribution to the scattering amplitude has quite a distinct conditional maximum given that the energy-momentum law is satisfied. This maximum point may be found either using any numerical method or analytically within some approximation \cite{Sharf:2006it}. Then one may use Laplace's method \cite{NGdeBruijn} to calculate the integrals for the cross-sections. The main contribution to the cross-section integral is made by the neighborhood region of the maximum point; according to Laplace's method this region is determined by the second derivatives of the logarithm of the integrand at the maximum point. This region is also very different from the multi-Regge one, since it does not force the produced particles to be strongly ordered in rapidity. In particular, when the energy $\sqrt{s}$ is close to the threshold energy for production of the $n$ secondary particles, all the rapidities at the maximum point are approximately equal and close to zero; moreover, the distances between the rapidities at the maximum point increase logarithmically slowly \cite{Sharph:2011wm} as the energy $\sqrt{s}$ grows, and decrease as the $n$ grows. Consequently, the strong ordering of the rapidities for a given multiplicity $n$ may appear only at the energies $\sqrt{s}$ such that the partial cross-section of the process with production of the $n$ particles is almost zero.

An absence of the strong ordering of rapidities makes it pointless to consider the multipheriperal diagrams without considering the interference contributions (which represent the various ways of joining of the identical particle lines to the diagram \cite{Sharf:2009yy,Sharph:2011wm}). Although the values of the interference contributions may be small, their number is too large to be neglected. Moreover, it has been shown (see \cite{ Sharf:2009yy}) that the contribution from the \textit{ladder} diagram is much smaller compared to the sum of the other contributions.

 The problem of an accounting for the interference contribution is considered in the paper \cite{InterferenceTerms} and briefly in \cite{TowPhysRevD.2.154}. However, in these papers the interference contributions have been considered within the approximations which are used in the ABFST model (see \cite{Amati1962}). Note that the assumption about the strong ordering of rapidities is used implicitly in the ABFST model (in order to simplify the integration limits in the integral equation for the $n$-particle contribution to the imaginary part of the scattering amplitude at zero four-momentum transfer). In addition, in the paper \cite{InterferenceTerms} a calculation of the cross-sections is considered only for the cases of the small multiplicities of produced particles, where the simple direct summation of the all interference contributions is possible. However, the number of the interference contributions increases rapidly (as $n!$) as the number of the secondary particles $n$ grows. Each interference contribution can be calculated quite easily with Laplace's method \cite{Sharf:2006it, Sharf:2007cf, Sharf:2009yy, Sharf:2011zz, Sharph:2011wm, Sharf:2013ona} even for the large number of the secondary particles $n$. However, it is problematic to account the \textit{large number} of such contributions even for the numerical computations. Therefore, the objective of this work is to propose an approximate calculation method which would allow one to calculate the sum of all the interference contributions for the processes with a large number of secondary particles ($n \sim 50$).
 
 The Bose-Einstein correlations \cite{Goldhaber:1959mj}, or the correlations between the four-momenta of the identity particles \cite{PhysRevLett.105.032001, CMS:2018vzv}, is almost the only interference effect discussed in the literature describing the experiment. This effect is typically considered in the context of the experimental determination of the dimension of the secondary particles radiation region \cite{doi:10.1142/S0217751X8900114X,Alexander_2003,Schegelsky2019}. A solution to this problem does not require a detailed description of the dynamics of inelastic process \cite{Khoze2016, CSORGO200015}. At the same time, it is the dynamics that makes the major difference between our model and the models \cite{Goldhaber:1960sf,GyulassyPhysRevC.20.2267,ZajcPhysRevD.35.3396,Weiner:1999th,WIEDEMANN1999145, KozlovUtyuzhWilkPhysRevC.68.024901,Utyuzh:2007ct,Kozlov_2002} which have been used for description of the interference effects. Let us discuss this difference.

The main problem when accounting for the interference contributions within the proposed model is that the real-valued integrands of different interference contributions reach their maximum at different points. If all the interference contributions had a common single maximum point, the sum of these interference contributions would also have a single maximum at that point. In this case one could use Laplace's method to calculate the whole sum of the interference contributions rather than calculating each contribution separately. Moreover, as we show below, in this case the maximum of such sum would become more \textit{pronounced} as $n$ grows. As a result, the accounting for the interference contributions would not be a problem for an arbitrarily large number of secondary particles. Meanwhile, in our model the different interference contributions have the different maximum points. Consequently, the sum of all interference contributions may have multiple maximum points depending on the energy $\sqrt{s}$ and the number of secondary particles $n$. The number of these maximum points may become so large that the time required for the numerical calculations increases critically. 
Even when the sum of all interference contributions have only a few maxima, they can be not pronounced enough under certain $\sqrt{s}$ and $n$ to limit to the quadratic terms in Laplace's method. 
It means that one should take into account the higher order terms in the Taylor expansion of the integrand logarithm around the maximum point. 
By contrast, the models considered in the papers \cite{Goldhaber:1960sf,GyulassyPhysRevC.20.2267,ZajcPhysRevD.35.3396,PhysRevLett.72.816,PhysRevC.50.469,PRATT1993159,Weiner:1999th,WIEDEMANN1999145, ALTINOLUK2016113} assume that the secondary particles are produced by the independent sources, and the Bose-Einstein correlations occur only when the particles propagate from the source to the detector. In this case different interference contributions are the Fourier transforms of the same function but at different points. 
As a result, the different interference contributions reach the maximum at the same point, which means that the dynamics in these papers is fundamentally different from the dynamics considered herein. The same is true for the models \cite{GyulassyPhysRevC.20.2267,WIEDEMANN1999145, KozlovUtyuzhWilkPhysRevC.68.024901,Kozlov_2002,ALTINOLUK2016113}, in which the problem of the field interacting with the fixed or random (Langevin) source is considered instead of the dynamics of interacting fields. In this case the secondary particle momentum distribution for an arbitrary number $n$ of the secondary particles is determined by the single function (i.e. by the Fourier transform of the external source); thus the different interference contributions reach the maximum at the same point. Such models also have an obvious problem associated with the violation of the energy-momentum conservation law caused by the space-time translation symmetry breaking. The latter is evidenced by the fact that, according to these models, as well as to the models based on the multi-Regge kinematics \cite{PhysRevD.95.034005}, the multiplicity of secondary particles has a Poisson distribution \cite{GyulassyPhysRevC.20.2267}; thus the particle production processes \cite{book:14428} at different regions of the phase space are independent of each other; consequently, the probability of the production of an arbitrarily large number of secondary particles is not zero. In addition, the models with an external source imply the source averaging with the corresponding density matrix \cite{PhysRevD.8.2284,PhysRevD.9.813,FOWLER1979349,PhysRevD.38.2209,Weiner:1999th}. However, the density matrix is actually postulated rather than determined with the time evolution operator obtained within some dynamical model.

From the above one may conclude that interference effects in the proton-proton scattering must be evident not only in the measurement of the Bose-Einstein correlations, but in any measured quantity.
In this paper we argue that the 
dependence of the inclusive rapidity distribution shape \cite{Berger:1972zh,Golokhvastov}, \cite{Thome:1977ky,BREAKSTONE1983458,GIACOMELLI,Alner:1986xu},\cite{ParticlesDataGroupPhysRevD.98.030001}(p. 590) on energy may be considered as an interference effect.
 where the shape of the inclusive rapidity distribution \cite{Berger:1972zh,Golokhvastov}, \cite{Thome:1977ky,BREAKSTONE1983458,GIACOMELLI,Alner:1986xu},\cite{ParticlesDataGroupPhysRevD.98.030001}(p. 590) depends on energy $\sqrt{s}$, may be considered as the interference effect. 
 The experimental data for the inclusive rapidity distribution is described by the different models \cite{Kaidalov:1983vn,Kaidalov:1983ew,WILK2007,NAVARRA2004568,Wolschin:2011mz,Werner:2010ny,PhysRevD.82.114006,Dash_2010,PhysRevC.83.044915,Jiang_2015} which however do not take into account for the interference effects. 
 The manifestation of the interference effects in the inclusive pseudorapidity distribution is considered in the papers \cite{PhysRevC.50.469,WIEDEMANN1999145}, though using the above mentioned approximations in which the different interference contributions reach maximum value at the same point. By contrast, in this paper we show that the behavior of the inclusive rapidity (pseudorapidity) distribution shape under the energy $\sqrt{s}$ growth is associated with the change of the distances between the maximum points of the different interference contributions. The mentioned difference between the dynamics considered in this paper and the one considered in papers \cite{PhysRevC.50.469,WIEDEMANN1999145} becomes apparent from the comparison of the shapes of inclusive rapidity distributions obtained within these models.
 
In order to focus on the problem of accounting for the interference contributions let us start with the simplest model $\phi^3$.

\section{Laplace's method, $\phi^3$ model}
\label{sec:phi3_and_definitions}

We consider the so-called \textit{ladder} diagrams for the inelastic $pp$ scattering within the real scalar $\phi^3$ model, assuming that the masses of the secondary particles are equal to the pion mass $m_{\pi}$. All the physical quantities used in this work (four-momenta, masses, energies...) are nondimensionalized by the pion mass $m_{\pi}$.

\begin{figure}[h]
\begin{center}
\includegraphics[width=1.0\linewidth]{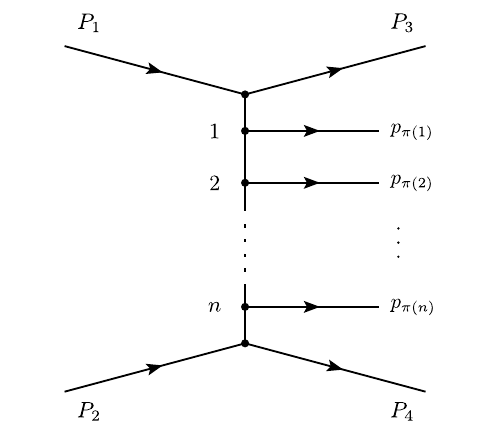}
\end{center}
\caption{Feynman ladder diagram within the $\phi^3$ model for inelastic $2 \rightarrow 2 + n$ scattering process. The diagram is associated with a permutation $\pi \in S_n$.}
\label{fig:ladder_diagram}
\end{figure}

 Let $S_n$ be the set of all possible permutations of $\mathbb{N}_n = \lbrace{1, 2, \dots, n\rbrace}$. Each diagram of the form \cref{fig:ladder_diagram} is characterized by a corresponding permutation $\pi \in S_n$. Assuming the one-line notation for the permutations, $\pi\left(i\right)$ denotes the index of the secondary particle which is joined to the $i$-th vertex in the corresponding diagram.
 
  Each permutation $\pi \in S_n$ specifies the diagram of the form \cref{fig:ladder_diagram}, and thus represents an analytic expression -- the additive contribution to the scattering amplitude
\begin{eqnarray}
a_n\left(P_1, P_2, P_3,P_4,p_{\pi(1)},...p_{\pi(n)}\right),
\label{single_a}
\end{eqnarray}
\noindent 
where $P_1, P_2$ are the four-momenta of the colliding particles in the initial state, $P_3, P_4$ are the four-momenta of the scattered particles in the final state, $p_{\pi(1)},...p_{\pi(n)}$ are the four-momenta of the secondary particles in the final state.  According to the Feynman diagram technique, the scattering amplitude $T_n$ of the considered process is the sum of all expressions $a_n$ corresponding to the elements of $S_n$
\begin{eqnarray}
\begin{aligned}
  T_n&\left(P_1, P_2, P_3,P_4,p_1,...p_n\right) = \\ &=
  \sum_{\pi \in S_n}
 a_n\left(P_1, P_2, P_3,P_4,p_{\pi\left(1\right)},...p_{\pi\left(n\right)}\right).
\end{aligned}
\label{amplitude}
\end{eqnarray}
For convenience, we introduce the following notations. We will consider the scattering process in the center-of-mass reference frame using the right-hand coordinate system in which $z$ axis is codirectional with ${\mathbf P}_1$. Here the expressions $a_n\left(P_1, P_2, P_3,P_4,p_{i_1},...p_{i_n}\right)$ depend on the real particles four-momenta which satisfy the 
following equations
%
\begin{eqnarray}
  \begin{aligned}
  & \left(P_i\right)_0^2 = M_i^2 + {\bm P}_i^2  \\
  & \left(p_i\right)_0^2 = m^2 + {\bm p}_i^2,
  \end{aligned}
\label{mass_surf_eq}
\end{eqnarray}
\noindent
where $M_i$ is the mass of the $i$-th incident particle in the initial state, and $m$ is the mass of the secondary particle.
The components of $P_1$ and $P_2$ in the chosen reference frame are $({\bm P}_1)_x = ({\bm P}_2)_x = 0$, $({\bm P}_1)_y = ({\bm P}_2)_y = 0$ and $({\bm P}_1)_z = -({\bm P}_2)_z$. Thus $P_1$ and $P_2$ can be uniquely determined by the single quantity $\sqrt{s} = (P_1)_0 + (P_2)_0$ -- the total energy of the colliding particles in the center-of-mass reference frame. We use the components $\left({\bm p}_i\right)_x$, $\left({\bm p}_i\right)_y$ of four-momenta and the rapidities $y_i$ of the secondary particles as the independent variables which uniquely determine all the four-momenta $p_i$. The rapidities $y_i$ are determined from the following expression 
\begin{eqnarray}
  \begin{aligned}
  (\bm p_i)_z = \sqrt{m^2 + ({\bm p}_i)_x^2 + ({\bm p}_i)_y^2}\sinh{\left(y_i\right)},
  \end{aligned}
\label{rapidities_def}
\end{eqnarray}
\noindent
 We also introduce another two independent variables $P^a_x$ and $P^a_y$ using the following expressions
 \begin{eqnarray}
  \begin{aligned}
  & P^a_x = \left({\bm P}_3\right)_x - \left({\bm P}_4\right)_x  \\
  & P^a_y = \left({\bm P}_3\right)_y - \left({\bm P}_4\right)_y.
  \end{aligned}
\label{p_a_xy}
\end{eqnarray}
 It follows from the equations \bref{mass_surf_eq} and the energy-momentum conservation law that all the components of $P_3$ and $P_4$ can be uniquely determined by specifying the $\sqrt{s}$, $P^a_x$, $P^a_y$ and all the $y_i$, $\left({\bm p}_i\right)_x$, $\left({\bm p}_i\right)_y$.

 Finally, instead of $4\left(n + 4\right)$ variables (the arguments of $a_n$) we have $3n + 3$ independent variables, so the expression \bref{single_a} may be rewritten as
\begin{eqnarray}
  \begin{aligned}
   a_n \left( \vphantom{\left({\bm p}_n\right)_y} \right. 
             \sqrt{s}, 
              y_1, \dots, y_n,
              \left({\bm p}_1\right)_x, \dots, \left({\bm p}_n\right)_x, \\ \left.
              \left({\bm p}_1\right)_y, \dots, \left({\bm p}_n\right)_y,
              P^a_x, P^a_y
      \right).
  \end{aligned}
\label{a_n_new_notations}
\end{eqnarray} 
 It is convenient to consider the introduced variables (except of $\sqrt{s}$) as the components of the some vector $X \in \mathbb{R}^{3n + 2}$.
\begin{eqnarray}
  \begin{aligned}
  X = \left( \vphantom{\left({\bm p}_n\right)_y} \right. 
           & y_1, \dots, y_n,
          \left({\bm p}_1\right)_x, \dots, \left({\bm p}_n\right)_x, \\ 
         & \left. \left({\bm p}_1\right)_y, \dots, \left({\bm p}_n\right)_y,
          P^a_x, P^a_y
      \right)
  \end{aligned}
\label{x_notation}
\end{eqnarray} 
Finally, it is also natural to introduce the notation for permutations of the components of $X$. To this end, to each permutation $\pi \in S_n$ we assign a linear operator $\hat{\pi} : \mathbb{R}^{3n + 2} \to \mathbb{R}^{3n+2}$ which can be defined as
\begin{eqnarray}
  \begin{aligned}
   \hat{\pi}X = X'= \left(\vphantom{X_i}\right.
         &X_{\pi(1)}, \dots, X_{\pi(n)}, \\
         & X_{n + \pi(1)}, \dots, X_{n + \pi(n)}, \\ 
         & X_{2n + \pi(1)}, \dots, X_{2n + \pi(n)}, \\ 
         &X_{3n+1}, \left. X_{3n+2}
      \right),
  \end{aligned}
\label{pi_hat_notation}
\end{eqnarray} 
\noindent
or the same: $\left(\hat{\pi}X\right)_{in + j} = X_{in + \pi(j)}$ for $i = 0..2$ and $j = 1..n$, while $\left(\hat{\pi}X\right)_i = X_i$ for $i = 3n+1, \ 3n+2$. One can see that according to the definition \bref{pi_hat_notation} of $\hat{\pi}$, we associate the operator $\hat{\pi}_2 \hat{\pi}_1$ with the composition of permutations $\pi_1\circ\pi_2$ such that
\begin{eqnarray}
  \begin{aligned}
  &\left( \hat{\pi}_2 \hat{\pi}_1 X \right)_{in + j} = 
  \left( \hat{\pi}_1 X \right)_{in + \pi_2\left(j\right)}
  = \\  & =
 \left( X \right)_{in + \pi_1\left(\pi_2\left(j\right)\right)} =
  \left( X \right)_{in + \left(\pi_1\circ\pi_2\right)\left(j\right)},
  \end{aligned}
\label{permutations_composition}
\end{eqnarray} 
\noindent
where $\hat{\pi}_i$ is the operator associated with $\pi_i$. Hence, using the introduced notations, the expression \bref{amplitude} can be rewritten in the following way
\begin{eqnarray}
 T_n\left( \sqrt{s}, X \right) =
 \sum_{\pi \in S_n}
 a_n\left( \sqrt{s}, \hat{\pi} X \right).
\label{amp_as_sum}
\end{eqnarray}
Now let us consider the expression for the partial cross-sections $\sigma_n$ of inelastic scattering, and rewrite this expression using the notations introduced above.
\begin{eqnarray}
\begin{aligned}
\sigma_n=
&\frac{\left(2\pi\right)^4}{4n!I}
\int
\frac{d{\mathbf P}_3}{(2\pi)^32{\left(P_3\right)}_0}
\frac{d{\mathbf P}_4}{(2\pi)^32{\left(P_4\right)}_0} 
\prod_{i = 1}^{n}\frac{d{\mathbf p}_i}{(2\pi)^32{\left(p_i\right)}_0}
\times \\ &\times  
\delta^4\left(P_1 + P_2 - P_3 - P_4 - \sum_i p_i\right)
\times \\ &\times  
\vert T_n \left(P_1,P_2,P_3,P_4,p_1,\dots, p_n \right)  \vert ^2,
\end{aligned}
\label{cs_origin}
\end{eqnarray}
\noindent
where $I = \sqrt{\left(P_1P_2\right)^2 - \left(M_1\right)^2 \left(M_2\right)^2 }$. First, we need to integrate out the energy-momentum conserving delta function in \bref{cs_origin} and then change the variables of integration. The Jacobian of the transformation and the multipliers appearing as the result of delta function integration can be included in the scattering amplitude $T_n$. This procedure described in detail in \cite{Sharph:2011wm}. As a result, the integral \bref{cs_origin} takes the following form
\begin{eqnarray}
\begin{aligned}
\sigma_n \left(\sqrt{s}\right) =
R\left(n, \sqrt{s}\right)
\int 
\prod_{i=1}^{3n+2}dX_i
\vert T_n \left( \sqrt{s}, X \right) \vert ^2, \ \ \
\end{aligned}
\label{cs_new_notations}
\end{eqnarray}
\noindent
where $R\left(n, \sqrt{s}\right) = \left(2\pi\right)^4/4n!I$.

Now, let us consider the problem of calculations of multidimensional integrals of the form \bref{cs_new_notations}.

\subsection{Calculation of the partial cross-sections within $\phi^3$ model}
\label{sec:sigma_calc}

\begin{figure}[h]
\center{\includegraphics[width=1\linewidth]{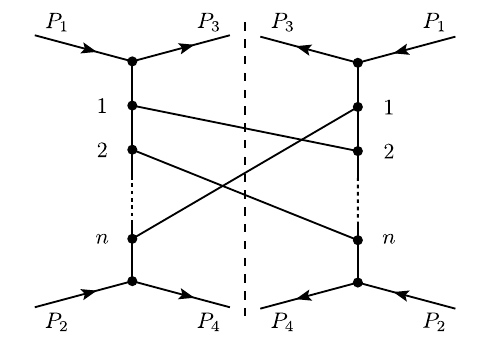}}
\caption{Cut diagram representing the interference contribution associated with particular permutation $\pi$ such that $\pi(1) = n$, $\pi(2) = 1$,..., $\pi(n) = 2$.}
\label{diagram_cut}
\end{figure}

Let us substitute the expression \bref{amp_as_sum} for the scattering amplitude $T_n\left(\sqrt{s}, X\right)$ in the expression of the partial cross-section \bref{cs_new_notations}.
\begin{eqnarray}
\begin{aligned}
\sigma_n & \left(\sqrt{s}\right) =
R\left(n, \sqrt{s}\right)
\times \\ \times
&\int 
\prod_{i=1}^{3n+2} dX_i
\sum_{\pi_k \in S_n} \sum_{\pi_l \in S_n}
\left[
a^*_n\left( \sqrt{s}, \hat{\pi}_k X \right)
 \right. \times \\ & \left. \times
a_n\left( \sqrt{s}, \hat{\pi}_l X  \right)
\right]
\end{aligned}
\label{cs_expanded_new_notations}
\end{eqnarray}
Note that  one of the two sums over $S_n$ in \bref{cs_expanded_new_notations} can be calculated by renaming the variables of integration.
\begin{eqnarray}
\begin{aligned}
\sigma_n & \left(\sqrt{s}\right) =
R\left(n, \sqrt{s}\right)
n! 
\times \\ \times
&\sum_{\pi \in S_n}
\int 
\prod_{i=1}^{3n+2} dX_i
a^*_n\left( \sqrt{s}, X \right)
a_n\left( \sqrt{s},\hat{\pi} X \right)
\end{aligned}
\label{cs_expanded_one_sum}
\end{eqnarray}  
The summands in \bref{cs_expanded_one_sum} are called the \textit{interference contributions} to the cross-section.  Each particular permutation $\pi$ specifies some interference contribution which can be represented by the so-called \textit{cut diagram} (\cref{diagram_cut}). The way in which the vertices from the both parts of a cut diagram are linked is specified by the permutation $\pi$ -- according to \bref{cs_expanded_one_sum} the $\pi\left(i\right)$ is the index of vertex in the left part of the diagram which is linked with the $i$-th vertex in the right part of the diagram.
 Each particular interference contribution can be calculated using the Laplace's method (see \cite{Sharph:2011wm}). This method allows one to easily calculate a multidimensional integral provided its integrand has a single maximum point.

Indeed, it has been shown (see \cite{Sharph:2011wm}) that the function $a_n\left(\sqrt{s}, X \right)$ has the single maximum point $X^{(0)}$ at fixed $\sqrt{s}$.
\begin{eqnarray}
  \begin{aligned}
  X^{(0)} = \left( 
           y^{(0)}_1, \dots, y^{(0)}_n,
          \left({\bm p}^{(0)}_1\right)_x, \dots, \left({\bm p}^{(0)}_n\right)_x,  \right. \\ \left.
          \left({\bm p}^{(0)}_1\right)_y, \dots, \left({\bm p}^{(0)}_n\right)_y,
          P^{a{(0)}}_x, P^{a{(0)}}_y
      \right)
  \end{aligned}
\label{max_point}
\end{eqnarray} 
\noindent
At the same time the function $a_n\left( \sqrt{s}, \hat{\pi} X \right)$ has the single maximum point $\hat{\pi}^{-1}X^{(0)}$, 
where $\hat{\pi}^{-1}$ is the operator \bref{pi_hat_notation} associated with permutation $\pi^{-1} \in S_n$ inverse of $\pi$; therefore the product $a^*_n\left( \sqrt{s}, X \right) a_n\left( \sqrt{s}, \hat{\pi} X \right)$ also has a single maximum point.
As a result, the Laplace's method can be applied to the calculation of the integrals in expression \bref{cs_expanded_one_sum}.

The calculation of the total cross-sections with Laplace's method even within the simplest $\phi^3$ model allows one to obtain the qualitative description of the experimental data (see \cite{Sharf:2006it, Sharf:2007cf, Sharf:2009yy, Sharf:2011zz, Sharph:2011wm, Sharf:2013ona}). Moreover, we plan to use Laplace's method to calculate the cross-sections within the multiparticle fields approach which is based on QCD (see \cite{Volkotrub:2015laa, MPFSS_2019}).

However, the number of the summands in \bref{cs_expanded_one_sum} is $n!$ which increases dramatically as the energy grows. It becomes impossible to account all of them by calculating each interference contribution separately. So we propose the method which makes it possible to account all the interference contributions for the processes with up to 50 secondary particles.


\section{The main idea of the proposed method }
\label{section_interference_contribs}

Let us consider a simple example to illustrate the main idea of the proposed method. We say that the maximum points of two functions, each having a single maximum point, are \textit{close} if the sum of these functions also has a single maximum point. Let $\phi_a\left(x\right) = exp\left(-\left(x - a\right)^2\right)$ be the function parametrized by the parameter $a$ and has single maximum point $x = a$. Now let us consider the sum
\begin{eqnarray}
g_a\left(x\right) = \phi_{-a}\left(x\right) + \phi_{-a/2}\left(x\right)+ \phi_{a/2}\left(x\right) + \phi_{a}\left(x\right)
\ \ \
\label{gfunc}
\end{eqnarray}
\noindent
 for different values of $a$, and the partial sums $\psi_-(x) = \phi_{-a}\left(x\right) + \phi_{-a/2}\left(x\right)$ and $\psi_+(x) = \phi_{a/2}\left(x\right) + \phi_{a}\left(x\right)$. It is easy to see that in the case of $a = 0$ the function $g_0\left(x\right)$ has the single maximum point $x = 0$. 
\begin{figure}[h]
\includegraphics[width=1.0\linewidth]{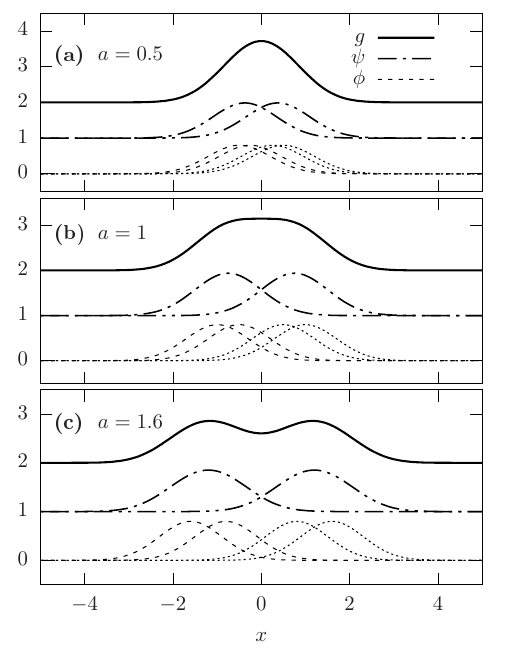}
\caption{Plot of the sum $g_a(x)$ for the different values of parameter $a$: (a) a = 0.5, (b) a = 1.0, (c) a = 1.6. The summands $\phi_a(x)$ are represented by the dashed lines, the partial sums $\psi_-(x)$ and $\psi_+(x)$ by dot-dashed lines and the total sum $g_a(x)$ by the solid line. Note that the functions are normalized and shifted vertically to avoid overlapping.}
\label{fig:func_example}
\end{figure}
Then, the distances between the maximum points of summands of $g_a(x)$ grow with the parameter $a$. For the values of the parameter $a$ up to $1$ each of the functions $g_a(x)$, $\psi_-(x)$ and $\psi_+(x)$ has a single maximum point, as can be seen in \cref{fig:func_example}.a and \cref{fig:func_example}.b. With further growth of $a$ the distances between the maximum points of the summands $\phi$ of $g_a(x)$ become so large that the single maximum of the function $g_a(x)$ \textit{splits} into two separate maxima (\cref{fig:func_example}.c). At the same time, the partial sums $\psi_-(x)$ and $\psi_+(x)$  still have the single maximum points.
   
The similar effect may be observed for the scattering amplitude $T_n\left(\sqrt{s}, X\right)$ with the energy growth at fixed $n > 1$. The function $T_n\left(\sqrt{s}, X\right)$ has a single maximum at low energies which then splits into a few separate maxima as the energy grows. To show this, let us consider the features of the maximum point $X^{(0)}$ of the function $a_n\left(\sqrt{s},X\right)$. At this point the transverse components of the particles momenta are equal to zero: $({\bm p}^{(0)}_i)_x = 0, ({\bm p}^{(0)}_i)_y = 0$, $P^a_x = 0$, $P^a_y = 0$, while the rapidities of the secondary particles form the following arithmetic progression (see \cite{Sharph:2011wm})
\begin{eqnarray}
\begin{aligned}
&y^{(0)}_i = y^{(0)}_1 - \Delta_y\left(i - 1\right) \text{, where } 
\\
&y^{(0)}_1 = \frac{n - 1}{2}\Delta_y, 
\\
&\Delta_y = \frac{2}{n + 1} \arcosh\left(\frac{\sqrt{s} - n}{2M}\right) 
\end{aligned}
\label{rap_progression}
\end{eqnarray}    
\noindent
As seen from \bref{rap_progression}, $y_i = - y_{n-i}$. Now consider the expression for the scattering amplitude $T_n\left(\sqrt{s}, X\right)$ i.e. the sum \bref{amp_as_sum} with taking into account the mentioned features of $X^{(0)}$. As we have already shown in the previous section, each summand of this sum (i.e. the function  $a_n\left(\sqrt{s},\hat{\pi}X\right)$) is associated with some permutation $\pi$ and reaches its maximum at the single point $\hat{\pi}^{-1}X^{(0)}$. The distance $\rho$ between the maximum points of two such summands (associated with $\pi_1$ and $\pi_2$) is defined by the following expression
\begin{eqnarray}
\begin{aligned}
\rho \left( \hat{\pi}^{-1}_1 X^{(0)},  \hat{\pi}^{-1}_2 X^{(0)}\right)& 
= \\ =
|\Delta_y| & \sqrt{\sum_{i=1}^{n}\left[ \pi^{-1}_1(i) - \pi^{-1}_2(i) \right]^2}. \ \ \ \
\end{aligned}
\label{max_point_distance}
\end{eqnarray}  
\noindent
Here we have taken into account that the $X_i = 0$  for $n < i < 3n + 2$. Given the fixed number of the secondary particles $n$, one can see from the \bref{max_point_distance} that the common difference $|\Delta_y|$ of the rapidities arithmetic progression  tends to zero as the energy $\sqrt{s}$ goes to the threshold value $2M + n$. In this case all the summands are \textit{close} so the whole sum \bref{amp_as_sum} has a single maximum point;  it means that one can apply Laplace's method to calculate the integral \bref{cs_new_notations}, thus taking into account all the interference contributions in the simple way. 

 However, the $|\Delta_y|$ slowly increases as the energy grows; and at some point the distances between the summands in \bref{amp_as_sum} become so large that the whole sum acquires several maxima. In other words, the single maximum of the scattering amplitude $T_n\left(\sqrt{s}, X\right)$ splits into a few separate maxima as the energy grows. We have already faced the similar effect for the one-dimensional function in the beginning of this section. In this case Laplace's method cannot be applied to the calculation of the integral \bref{cs_new_notations} since its integrand has several maxima. One could still consider the sum \bref{cs_expanded_one_sum} and calculate each interference contribution with Laplace's method, but this way it is impossible to calculate all the interference contributions for the processes with a large number of secondary particles.

  \textit{The main idea of the proposed method} of accounting for the interference contributions is to apply Laplace's method not to the each interference contribution separately, but to the sums of the interference contributions whose integrands have the \textit{close} maximum points. The integrand of such sum has a single maximum point so the whole sum can be calculated with Laplace's method. Although the scattering amplitude $T_n\left(\sqrt{s}, X\right)$ acquires several maxima as the energy grows, we can split the sum \bref{amp_as_sum} into subsums each having a single maximum point. To this end we group the neighboring vertices of the diagram into $k$ groups. Then we group all the permutations $\pi \in S_n$ into subsets $I_1, I_2, \dots$ in such a way that any two permutations belong to the same subset if and only if they specify the diagrams in which the lines of the secondary particles are joined to the same groups of the vertices regardless of the order within the groups. So the diagrams specified by the permutations belonging to the same subset $I_i$ differ only in the order of joining inside the groups of vertices. Therefore, according to \bref{rap_progression} and \bref{max_point_distance}, we expect the functions associated with such diagrams to have the \textit{close} maximum points. Thus, for each subset $I_i$ we consider the sum $b_i \left(\sqrt{s}, X\right) = \sum_{\pi \in I_i}a_n\left(\sqrt{s}, \hat{\pi}X\right)$. For any given values of $n$ and $\sqrt{s}$ we select such a number and the sizes of the groups that make each of the functions $b_1$, $b_2$, $\dots$ have a single maximum point. As a result, one can use Laplace's method to calculate the sum of the interference contributions associated with each subset $I_i \subset S_n$ rather than calculating the interference contribution associated with each permutation $\pi \in S_n$. 

Such approach reduces greatly the amount of the calculations and allows one to account for all the interference contributions for the processes with the high production of the secondary particles (up to 50 secondary particles at 72 GeV at present). The specific implementation of the proposed idea will be described in the next section in detail.

\subsection{Grouping the vertices of the diagram}
\label{subsection_groups_partition}

Let us group the neighboring vertices of the diagram \cref{fig:ladder_diagram} into the $k$ \textit{groups} (non-empty sets) each containing $n_i$ vertices.
\begin{figure}[h]
\center{\includegraphics[width=1\linewidth]{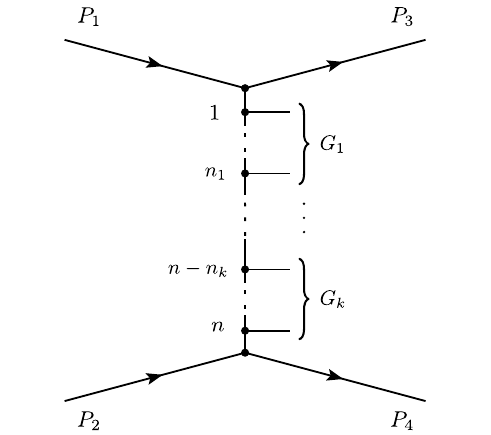}}
\caption{Grouped vertices.}
\label{pic_grouped_vertices}
\end{figure}
\noindent
For this purpose we consider the sequence of sets $G_1, G_2, \dots, G_k$ each specifying a corresponding group. The set $G_i$ contains the indices of vertices that are grouped into $i$-th group. 
\begin{eqnarray}
\begin{aligned}
G_i = \left\{ \nu \in \mathbb{N}_n \ \mid \ \sum_{k=1}^{i - 1} n_k < \nu \leqslant \sum_{k=1}^{i} n_k \right\}
\end{aligned}
\label{G_i_sets}
\end{eqnarray}  
\noindent
for $1 \leqslant i \leqslant k$, where $n_i$ is the number of vertices in the $i$-th group.

\begin{figure}[h]
\center{\includegraphics[width=1\linewidth]{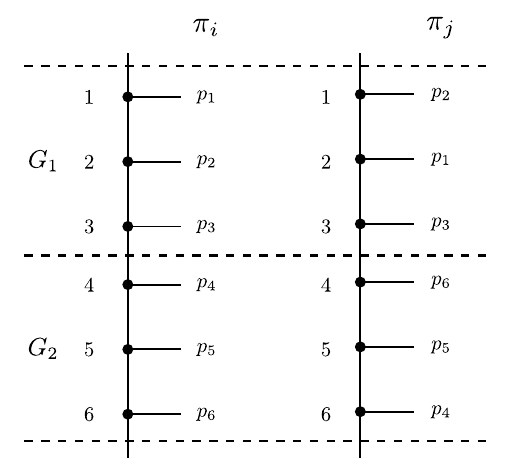}}
\caption{Example of equivalent permutations $\pi_i$ and $\pi_j$ given $n = 6$, $G_1 = \lbrace{1,2,3\rbrace}$ and $\ G_2 = \lbrace{4,5,6\rbrace}$. Here $\pi_i\left(G_1\right) =  \pi_j\left(G_1\right) = \lbrace{1,3,5\rbrace}$ and 
$\pi_i\left(G_2\right) =  \pi_j\left(G_2\right) = \lbrace{2,4,6\rbrace}$, which imply that $\pi_i \sim \pi_j$.}
\label{fig:equivalence_example}
\end{figure}

\begin{figure}[h]
\center{\includegraphics[width=1\linewidth]{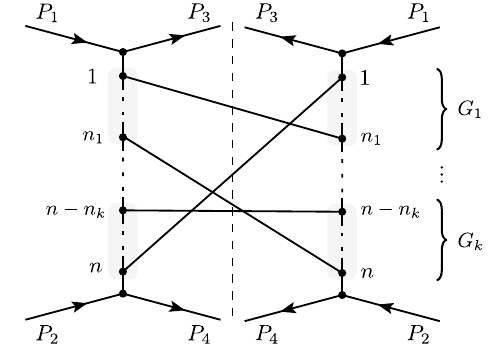}}
\caption{Cut diagram in which the vertices of the both parts are grouped into $k$ groups.}
\label{diagram_cut_grouped}
\end{figure}

 The lines of the secondary particles in the connected diagram specified by a permutation $\pi$ are distributed among the groups of the vertices in some way depending on the permutation. Thus for any permutation $\pi \in S_n$ we can also consider the sets $\pi\left(G_1\right), \pi\left(G_2\right), \dots, \pi\left(G_k\right)$, where $\pi\left(G_i\right) = \lbrace{\pi\left(\nu\right) \mid \nu \in G_i\rbrace}$ is the set containing the indices of the secondary particles which are joined to the vertices of the $i$-th group in the diagram specified by the permutation $\pi$. It allows us to introduce the equivalence relation $\sim$ on the permutations set $S_n$ in the following way
\begin{eqnarray}
\begin{aligned}
\pi_i \sim \pi_j
\
\Leftrightarrow
\
\forall l \in [1..k], \ \pi_i\left(G_l\right) = \pi_j\left(G_l\right)
\end{aligned}
\label{equivalence_relation}
\end{eqnarray} 
\noindent
Which means that any two permutations $\pi_i$ and $\pi_j$ are equivalent if they specify the diagrams in which the external lines of the secondary particles are joined to the same groups of the vertices regardless of the joining order inside the groups (\cref{fig:equivalence_example}).
As a result, the permutations set $S_n$ may be split into the equivalence classes $[\pi] = \left\{ \pi_i \in S_n \ \mid \ \pi_i \sim \pi \right\}$. In other words, the introduced equivalence relation $\sim$ provides the partition $S_n / _\sim$ of the underlying set $S_n$, where $S_n / _\sim$ is the quotient set of the permutations set $S_n$ by $\sim$ i.e. the set contains all the equivalence classes  $[\pi]$.
\begin{eqnarray}
S_n = \bigcup_{[\pi] \in {S_{n}}{/\sim}} [\pi]
\label{quotient_set}
\end{eqnarray} 
\noindent
Taking into account \bref{quotient_set}, the expression \bref{amp_as_sum} may be rewritten as
\begin{eqnarray}
T_n\left(\sqrt{s}, X\right) = \sum_{[\pi] \in {S_{n}}{/\sim}}
 \left[
 \sum_{\pi_i \in [\pi]} a_n\left( \sqrt{s},\hat{\pi}_i X \right)
 \right]
\label{amp_throug_quotient_set}
\end{eqnarray} 
\noindent
Now let us consider the auxiliary function $A_n\left(\sqrt{s}, X\right)$
\begin{eqnarray}
A_n\left(\sqrt{s}, X\right) =
 \sum_{e \in [\epsilon]}
  a_n\left( \sqrt{s},\hat{e} X \right)
\label{amp_internal_permut_sum}
\end{eqnarray} 
\noindent
where $[\epsilon]$ is the equivalence class containing the permutations equivalent to the \textit{identity} permutation $\epsilon$ such that $\epsilon \left(i\right) = i$ for all $ i \in [1 \dots n]$. The maximum distance between the maximum points of summands in \bref{amp_internal_permut_sum} may be significantly reduced compared to the summands of \bref{amp_as_sum} by increasing the number of groups $k$ and decreasing their sizes $n_1, n_2, \dots n_k$. Therefore, for any given values of $n$ and $\sqrt{s}$ one may select such a number of the groups $k$ and sizes of the groups $n_1, n_2, \dots n_k$ that make function $A_n\left(\sqrt{s}, X\right)$ have a single maximum point. 
%
 It may be shown that for any permutation $\pi_i \in [\pi]$ there exists a permutation $e \in [\epsilon]$ such that $\pi_i = \pi\circ e$, where $\pi$ is the representative (any member) of the class $[\pi]$; taking this into account and also \bref{permutations_composition}, let us represent the scattering amplitude \bref{amp_throug_quotient_set} in terms of $A_n\left(\sqrt{s}, X\right)$
\begin{eqnarray}
\begin{aligned}
T_n\left(\sqrt{s}, X\right) =
 \sum_{[\pi] \in {S_{n}}{/\sim}}
 \left[
 \sum_{e \in [\epsilon]} a_n\left( \sqrt{s},\hat{e} \left(\hat{\pi}X\right) \right)
 \right] &
 = \ \ \ \\ =
  \sum_{[\pi] \in {S_n / \sim}} A_n\left(\sqrt{s}, \hat{\pi} X\right) &
\end{aligned}
\label{T_through_A}
\end{eqnarray} 
\noindent
where $\hat{\pi}$ is the permutation operator \bref{pi_hat_notation} associated with the representative of $[\pi]$. As a result, the integral \bref{cs_new_notations} may also be written in terms of $A_n\left(\sqrt{s}, X\right)$.
\begin{eqnarray}
\begin{aligned}
\sigma_n \left(\sqrt{s}\right) =
R\left(n, \sqrt{s}\right)
\times \\ \times
\int 
\prod_{i=1}^{3n+2} dX_i
\left[
\sum_{[\pi_i] \in {S_n / \sim}}
A^*_n\left(\sqrt{s}, \hat{\pi}_i X\right) \right.
\times \\ \times \left.
\sum_{[\pi_k] \in {S_n / \sim}}
A_n\left(\sqrt{s}, \hat{\pi}_k X\right) \right]
\end{aligned}
\label{cs_through_A_double_sum}
\end{eqnarray}  
\noindent
One of the two summations in the integral \bref{cs_through_A_double_sum} may be calculated by renaming the variables of integration  as it was done between \bref{cs_expanded_new_notations} and \bref{cs_expanded_one_sum}. Note that in this case the number of the elements in the set of summation ${S_n / \sim}$ is $n!/n_1!n_2!\dots n_k!$. 
\begin{eqnarray}
\begin{aligned}
&\sigma_n  \left(\sqrt{s}\right) =
R\left(n, \sqrt{s}\right)
\frac{n!}{n_1!n_2!\dots n_k!}
\times \\ \times
&\sum_{[\pi] \in {S_n / \sim}}
\int 
\prod_{i=1}^{3n+2} dX_i
A^*_n\left(\sqrt{s},X\right)
A_n\left(\sqrt{s}, \hat{\pi} X\right)
\end{aligned}
\label{cs_through_A}
\end{eqnarray}
Taking into account that the function $A_n\left(\sqrt{s}, X\right)$ has a single maximum point, Laplace's method may be used to calculate the summands of \bref{cs_through_A}. The number of the summands in \bref{cs_through_A}, which is equal to  $\vert S_n/\sim \vert = n!/n_1!n_2!\cdots n_k!$, may be significantly reduced by selecting the number of group $k$ such that $k << n$. Each summand of \bref{cs_through_A} can also be represented by the corresponding cut diagram \cref{diagram_cut_grouped}.  
Hence, according to \bref{cs_through_A}, it is sufficient to consider only a single permutation from each equivalence class $[\pi]$ rather then all permutations of $S_n$ in order to calculate all the interference contributions to the cross-section $\sigma_n$. Moreover, we will show that there are many similar summands in \bref{cs_through_A} which may be easily calculated by the corresponding weight-factors.

 Let us consider a summand of \bref{cs_through_A} associated with a class $[\pi_i]$. Then change the integration variables: $X \to \hat{e}X$, where $e $ is a permutation belonging to the class $[\epsilon]$. Note that the function $A_n$ is symmetric with respect to such transition: $A_n\left(\sqrt{s},\hat{e}X\right) = A_n\left(\sqrt{s}, X\right)$.
\begin{eqnarray}
\begin{aligned}
&\int dX A^*_n\left(\sqrt{s},X\right) A_n\left(\sqrt{s}, \hat{\pi}_i X\right)
= \\ = 
&\int d\left(\hat{e}X\right) A^*_n\left(\sqrt{s},\hat{e}X\right) A_n\left(\sqrt{s}, \hat{\pi}_i \hat{e}X\right)
=  \\ =
&\int dX A^*_n\left(\sqrt{s},X\right) A_n\left(\sqrt{s}, \hat{\pi}_l X\right)
\end{aligned}
\label{diff_permuts_same_contrib}
\end{eqnarray}
\noindent
According to the definition of permutation operator \bref{pi_hat_notation}, the operator $\hat{\pi_l}=\hat{\pi}_i\hat{e}$ is assigned to the permutation $\pi_l=e\circ\pi_i$.
It may be shown that the permutations $\pi_i$ and $e$ may be selected such that $\pi_i$ and $\pi_l = e\circ\pi_i$ belong to the different equivalence classes $[\pi_i]$ and $[\pi_l]$.
As a result, one can see from \bref{diff_permuts_same_contrib} that there are two similar contributions to the sum \bref{cs_through_A} associated with the different equivalence classes.

In other words, the definition of the function $A_n$ implies that \textit{both} parts of the cut diagram \cref{diagram_cut_grouped} are symmetric with respect to the order of joining inside the groups of vertices. Consequently, the permutations which establish the same number of connections between the groups of vertices differ only in the order of linking inside the groups and therefore represent the similar contributions to the sum \bref{cs_through_A}.
To clarify this statement, let us consider the number of connections $m_{ij}$ between the $i$-th group in the left part and the $j$-th group in the right part of the cut diagram \bref{diagram_cut_grouped} specified by a permutation $\pi$.
\begin{eqnarray}
\begin{aligned}
 m_{ij} = \left\vert\pi\left(G_i\right)\cap G_j\right\vert
\end{aligned}
\label{matrix_elements}
\end{eqnarray}
\noindent
where $|\cdot|$ denotes the number of elements in the set (i.e. the power of the set). In this way, each equivalence class $[\pi]$ may be characterized by a matrix $m$ of size $k \times k$ whose elements are defined by \bref{matrix_elements}. If any two classes $[\pi_i]$ and $[\pi_j]$ are characterized by the same matrix $m$, then they specify the \textit{equivalent} cut diagrams, which mean that they correspond to the similar contributions to the sum \bref{cs_through_A} because one of these diagrams can be obtained from another by rearranging the vertices inside the groups.
\begin{figure}[h]
\center{\includegraphics[width=1\linewidth]{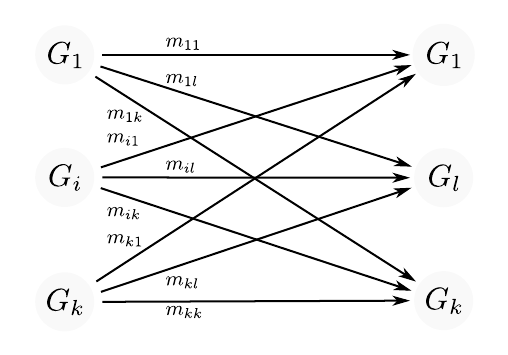}}
\caption{The number of the connections $m_{ij}$ between the groups in the cut diagram specified by some permutation $\pi$.}
\label{matrix_interpret}
\end{figure}

In order to consider all the unique summands of \bref{cs_through_A}, we consider the set $M$ of all possible matrices $m$ of the size $k \times k$ such that the sum of the elements of $i$-th row (column) is equal to the number of vertices in the $i$-th group.
\begin{eqnarray}
\begin{aligned}
	M = \left\{
	\vphantom{\sum}
m \in M_{k,k}\left(\mathbb{N}\right) 
\ : \ 
\sum_{j = 1}^{k}m_{ij} = \sum_{j = 1}^{k}m_{ji} = n_i
\right\} \ \ \
\end{aligned}
\label{group_matrix}
\end{eqnarray}
%
%
Each matrix $m$ represents a summand in \bref{cs_through_A}. All the summands similar to this one can be accounted by weight factor $W_m$ uniquely determined by the elements of the matrix $m$. To calculate the summand specified by the matrix $m$, one needs to link the vertices of the cut diagram in a way specified by the matrix (\cref{matrix_interpret}). As a result, one obtains the permutation $\pi_m$ such that $\left\vert\pi_m\left(G_i\right)\cap G_j\right\vert = m_{ij}$, which makes it possible to calculate  summand associated with the class $[\pi_m]$.
The weight factor $W_m$ may be calculated by counting the number of different equivalence classes $[\pi]$ characterized by the same matrix $m$. The number of such classes is equal to the number of all possible ways in which one can group the elements of the sets $G_1, G_2, \dots, G_n$ into $k$ sets $G_1', G_2',\dots, G_k'$ such that $\vert G_i \cap G_j'\vert = m_{ij}$, where $i, j \in [1..k]$; so that each set $G_j'$ contains exactly $m_{ij}$ elements from the set $G_i$.
\begin{eqnarray}
\begin{aligned}
W_m =
\prod_{i = 1}^{k}
\frac{n_i!}{\prod_{j = 1}^{k}m_{ij}!}
\end{aligned}
\label{weight_factor}
\end{eqnarray}
Finally, the expression for the partial cross-section may be rewritten as
\begin{eqnarray}
\begin{aligned}
&\sigma_n \left(\sqrt{s}\right) =
R\left(n, \sqrt{s}\right)
\frac{n!}{n_1!n_2!\dots n_k!}
\times \\ \times
&
\sum_{m \in M}
W_m 
\int 
\prod_{i=1}^{3n+2} dX_i
A^*_n\left(\sqrt{s},X\right)
A_n\left(\sqrt{s}, \hat{\pi}_m X\right)
\ \
\end{aligned}
\label{cs_final}
\end{eqnarray}
Note that the number of summands in \bref{cs_final} may be much smaller compared to the sum \bref{cs_origin}. At the same time, Laplace's method may be used to calculate each particular summand of the sum \bref{cs_final}.

It remains now to consider in detail all the components necessary to apply Laplace's method within the proposed approach for calculation of interference contributions. 

First we consider the feature of the maximum point of the function $A_n\left(\sqrt{s}, X \right)$. We have already mentioned the features of the maximum point of  $a_n\left(\sqrt{s}, X\right)$ in \cref{section_interference_contribs}. Since the function $a_n\left(\sqrt{s}, X\right)$ grows as the $\left({\mathbf p}_{\perp}\right)_i \to 0$ and ${\mathbf P}^a_\perp \to 0$, the function \bref{amp_internal_permut_sum} reaches the maximum value at a point $\chi^{(0)}$ where $\left({\mathbf p}_{\perp}\right)_i = 0, \ i=1..n$ and ${\mathbf P}^a_\perp = 0$; so $\chi^{(0)}_i = 0$ for $n < i \leqslant 3n+2$.

 As we mentioned above, we group the vertices of the diagram in a way for the function  $A_n\left(\sqrt{s}, X\right)$ to have a \textit{single} maximum point. Taking also into account that $A_n\left(\sqrt{s}, \hat{e}X\right) = A_n\left(\sqrt{s}, X\right)$ for any $e \in [\epsilon]$, one can conclude that
%
%
\begin{eqnarray}
\begin{aligned}
\hat{e}\chi^{(0)} = \chi^{(0)}
\end{aligned}
\label{chi0_feature}
\end{eqnarray}
\noindent
for any  $e \in [\epsilon]$. This important feature of $\chi^{(0)}$ allows one to calculate the value of function $A_n$ together with its $2$-nd derivatives at the maximum point in a simple way. 
\begin{eqnarray}
\begin{aligned}
 A_n^{0} = A_n\left(\sqrt{s}, \chi^{(0)} \right)
 =
\sum_{e \in [\epsilon]}
a_n\left(\sqrt{s}, \hat{e} \chi^{(0)} \right)&
 = \\ =
\left(\prod_{i = 1}^{k}n_i!\right)
a_n\left(\sqrt{s},\chi^{(0)} \right)&
\end{aligned}
\label{Amax}
\end{eqnarray}  
 \noindent
 Let us consider the first $n$ components of the vector of derivatives $\partial A_n / \partial X_i$ of $A_n$  at the maximum point $\chi^{(0)}$ assuming that $\chi^{(0)}$ satisfies the equation \bref{chi0_feature}. We denote these components by $\partial A_n / \partial y_i$
\begin{eqnarray}
\begin{aligned}
 \frac{\partial A_n}{\partial y_i}  \left(\sqrt{s}, \chi^{(0)} \right) 
 =
 \sum_{e \in [\epsilon]} \frac{\partial a_n}{\partial y_{e^{-1}(i)}}  \left( \sqrt{s},\hat{e} \chi^{(0)} \right)&
 = \\  =
 \frac{\prod_{j = 1}^{k}n_j!}{n_{g(i)}}
 \sum_{j \in G_{g(i)}} 
 \frac{\partial a_n}{\partial y_j}\left( \sqrt{s},\chi^{(0)} \right) &
\end{aligned}
\label{dA_dy}
\end{eqnarray} 
\noindent
where $g\left(i\right)$ is the index of the group containing the index $i$ (so that $i \in G_{g(i)}$), $e^{-1}$ is the inverse permutation of $e$, i.e. such that $e^{-1} \circ e = \epsilon$. The derivatives $\partial A_n / \partial \left(p_x\right)_i$ and $\partial A_n / \partial \left(p_y\right)_i$ are calculated in the same way as $\partial A_n / \partial y_i$. For the last two components of $\partial A_n / \partial X_i$ we have
\begin{eqnarray}
\frac{\partial A_n}{\partial X_i}  \left(\sqrt{s}, X\right) 
  = \prod_{j = 1}^{k}n_j!\frac{\partial a_n}{\partial X_i}\left( \sqrt{s}, X \right)
\label{dA_dPaXPaY}
\end{eqnarray} 
\noindent
where $i = 3n+1, 3n+2$.

 The matrix of the second derivatives $\partial^2 A_n / \partial X_i \partial X_j\left(\sqrt{s}, \chi^{(0)}\right)$ may be 
 split into blocks with respect to the variables: $y_i$ and $\left({\mathbf p}_x\right)_x$, $\left({\mathbf p}_x\right)_y$. Then each block may be considered separately. Let us start with the block containing the derivatives $\partial^2 A_n / \partial y_i \partial y_j\left(\sqrt{s}, \chi^{(0)}\right)$ which we denote briefly by $D^{(0)}_{ij}$.
\begin{eqnarray}
 D^{(0)}_{ij}
 =
 \sum_{e \in [\epsilon]} \frac{\partial^2  a_n}{\partial y_{e^{-1}(i)} \partial y_{e^{-1}(j)}}  \left( \sqrt{s},\hat{e} \chi^{(0)} \right)
\label{d2A_dy}
\end{eqnarray} 
\noindent
 In order to calculate the sum \bref{d2A_dy}, we denote the derivatives $\partial^2 a_n / \partial y_{i} \partial y_{j} \left( \sqrt{s}, \chi^{(0)} \right)$ by $d^{(0)}_{ij}$ and consider the following cases keeping in mind that $\chi^{(0)}$ satisfies the equations \bref{chi0_feature}:
\begin{enumerate}
\item{$g(i) = g(j) \ \text{and}\ i = j$
\begin{eqnarray}
 D^{(0)}_{ij}  =
  \frac{\prod_{l = 1}^{k}n_l!}{n_{g(i)}}
  \sum_{i_1 \in G_{g(i)}} 
   d^{(0)}_{i_1 i_1}
\label{d2A_dy1}
\end{eqnarray} 
}
\item{$ g(i) = g(j) \ \text{and}\ i \neq j$
\begin{eqnarray}
 D^{(0)}_{ij} =
  \frac{\prod_{l = 1}^{k}n_l!}{(n_{g(i)} - 1)n_{g(i)}}
  \sum_{\substack{i_1, i_2 \in G_{g(i)} \\ i_1 \neq i_2}} 
  d^{(0)}_{i_1 i_2}
\label{d2A_dy2}
\end{eqnarray} 
}
\item{$g(i) \neq g(j)$
\begin{eqnarray}
  D^{(0)}_{ij} =
  \frac{\prod_{l = 1}^{k}n_l!}{n_{g(i)}n_{g(j)}}
  \sum_{\substack{i_1 \in G_{g(i)} \\ i_2 \in G_{g(j)}}} 
  d^{(0)}_{i_1 i_2}
\label{d2A_dy3}
\end{eqnarray}
}
\end{enumerate}
\noindent
The rest of the blocks containing the derivatives $\partial^2 A_n / \partial \left({\mathbf p}_x\right)_i \partial \left({\mathbf p}_x\right)_j  \left(\sqrt{s}, \chi^{(0)}\right)$  and $\partial^2 A_n / \partial \left({\mathbf p}_y\right)_i \partial \left({\mathbf p}_y\right)_j  \left(\sqrt{s}, \chi^{(0)}\right)$ may be calculated in the same way as \bref{d2A_dy1} - \bref{d2A_dy3}. Note that according to \bref{d2A_dy1} - \bref{d2A_dy3}, all the mixed derivatives $\partial^2 A_n / \partial y_i \partial \left({\mathbf p}_x\right)_j  \left(\sqrt{s}, \chi^{(0)}\right)$ and $\partial^2 A_n / \partial y_i \partial \left({\mathbf p}_y\right)_j  \left(\sqrt{s}, \chi^{(0)}\right)$ are equal to zero, since the derivatives $\partial^2 a_n / \partial y_i \partial \left({\mathbf p}_x\right)_j  \left(\sqrt{s}, X^{(0)}\right)$ and $\partial^2 a_n / \partial y_i \partial \left({\mathbf p}_y\right)_j  \left(\sqrt{s}, X^{(0)}\right)$ are equal to zero (see \cite{Sharph:2011wm}). 

Thus, we obtain all the components necessary to calculate the sum \bref{cs_final} with Laplace's method.

\section{Results}

\begin{figure}[h]
\begin{center}
\includegraphics[width=1\linewidth]{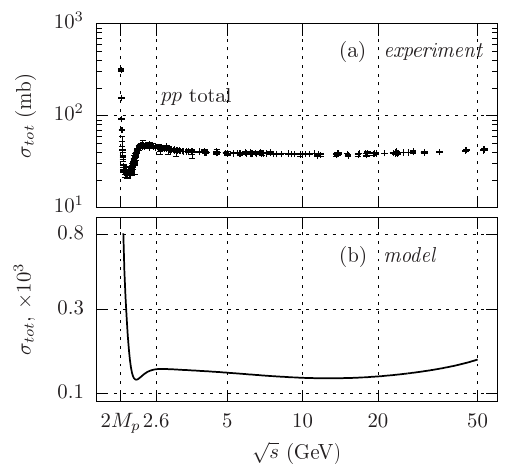}
\end{center}
\caption{The energy dependence of the total cross-section of $pp$ scattering. Experimental data \cite{Agashe:2014kda} (a) and the dependency calculated with the proposed method taking into account all the interference contributions for the digrams with up to $n = 50$ secondary particles (b). $M_p$ is the proton mass. }
\label{fig:total_cross-section}
\end{figure}

\begin{figure}[h]
\begin{center}
\begin{minipage}[h]{1\linewidth}
\includegraphics[width=1\linewidth]{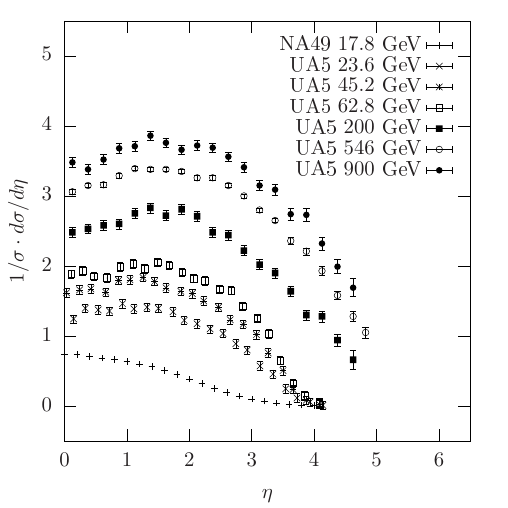}
\end{minipage}
\begin{minipage}[h]{1\linewidth}
\includegraphics[width=1\linewidth]{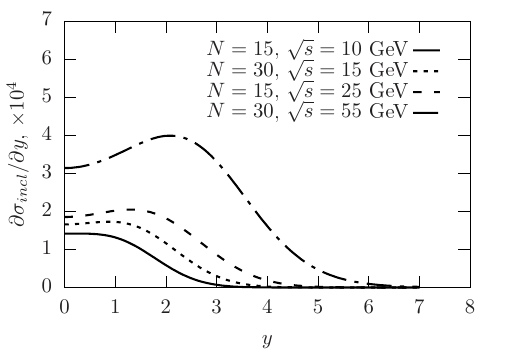}
\end{minipage}
\end{center}
\caption{Rapidity (pseudorapidity) distributions in $pp$ and $p\bar{p}$ collisions at various center of mass energies. Experimental data \cite{TheNA49Collaboration2006, Thome:1977ky} (a). Rapidity distribution calculated with the proposed method within $\phi^3$ model (b) and (d). Here $N$ denotes the maximum number of the secondary particles for which all the interference contributions are calculated.}
\label{fig:rapidity_distributions}
\end{figure}

We applied the proposed method within the $phi^3$ model to calculate the energy dependence of the $pp$ scattering total cross-section and the inclusive rapidity distribution for inelastic $pp$ scattering. The models presented in \cref{fig:total_cross-section} and \cref{fig:rapidity_distributions} are obtained at the value of effective coupling constant $L = 29$ (the expression for $L$ is the same as in \cite{Sharph:2011wm}, page 876).

 As we can see from \cref{fig:total_cross-section} and \cref{fig:rapidity_distributions}, comparison with the experimental data shows that the proposed approach allows us to obtain the theoretical predictions that are in qualitative agreement with the experimental data even within the simplest model.

\noindent
The following expression has been used for the calculations of inclusive rapidity distribution

\begin{eqnarray}
\begin{aligned}
\frac{\partial \sigma_{incl}}{\partial y} \left(\sqrt{s}, y\right)
\propto
\sum_{n = 1}^{N\left(\sqrt{s}\right)}
\sum_{i = 1}^{n}
\int 
\prod_{\substack{j = 1 \\ j \neq i}}^{3n+2} dX_j
\vert T_n \left( \sqrt{s}, X \right)\vert ^2 
\biggr\rvert_{X_i = y},
\end{aligned}
\label{dsigma_dy}
\end{eqnarray}

\noindent
where $N\left(\sqrt{s}\right)$ is for the maximum available number of the secondary particles (each having mass $m$) which may be produced as the result of inelastic scattering at a fixed energy value $\sqrt{s}$.

It is of interest to consider the peaks behavior in \cref{fig:rapidity_distributions}. At low energies one can observe the single peak at rapidities (pseudorapidities) close to zero. However, this peak becomes less pronounced and transforms into the so-called \textit{rapidity plateau} as the energy $\sqrt{s}$ grows. With further energy growth the plateau splits into the two separate peaks at non-zero points located symmetrically about zero; 

The effect described above may be explained by analyzing the behavior of the maximum points of the scattering amplitude absolute value $\vert T_n \left( \sqrt{s}, X \right) \vert^2$ with the energy growth. Note that the function $\vert T_n \left( \sqrt{s}, X \right) \vert$ has the same maximum points as $ T_n \left( \sqrt{s}, X \right)$ since the scattering amplitude \bref{amp_as_sum} within $\phi^3$ model is the real-valued function, except for the constant complex factor which may be neglected.

\begin{eqnarray}
T_n\left(\sqrt{s}, X\right) = \sum_{\pi \in S_n} a_n\left( \sqrt{s},\hat{\pi} X \right)
\label{amp_as_sum_again}
\end{eqnarray} 

\noindent

Let us consider \bref{amp_as_sum_again} at the values of energy close to the inelastic threshold energy $\sqrt{s} \gtrsim 2M + m$. In this case, according to \bref{rap_progression}, for any $n = 1,..,N\left(\sqrt{s}\right)$ the rapidities of the secondary particles $y^{(0)}_1, y^{(0)}_2, \dots, y^{(0)}_n$ at the maximum point of the function $a_n\left( \sqrt{s}, X \right)$ are close to zero. Thus, the summands in \bref{amp_as_sum_again} have \textit{close} maximum points so the function $T_n \left( \sqrt{s}, X \right)$ has the single maximum point $\zeta^{(0)}$. It means also that the single group ($k = 1$) is enough to make \bref{amp_internal_permut_sum} the function  $A_n \left( \sqrt{s}, X \right)$ ( which in this case is identically equal to  $T_n \left( \sqrt{s}, X \right)$) have a single maximum point. Taking this into account and also \bref{rap_progression}, \bref{chi0_feature} one may conclude that $\zeta^{(0)} = (0, \dots, 0)$. Hence, the most probable inelastic processes at low energies are the processes in which all the secondary particles have near-zero-rapidities. I.e. the mean value of the vector of secondary particle rapidities at low energies is equal to zero. As a result, the inclusive rapidity distribution has the single pronounced maximum at zero.

The continuous growth of the energy $\sqrt{s}$ increases the distances between the maximum points of the summands in \bref{amp_as_sum_again}. As a result, the single maximum of $T_n \left( \sqrt{s}, X \right)$ becomes \textit{less pronounced}, i.e. the absolute values of eigenvalues $|\lambda_i|$ of the Hessian matrix (with respect to $y$) of the function $T_n$ at the maximum point $\zeta^{(0)}$ decrease and tend to zero. It means that the variances of the rapidities of the secondary particles are increasing with energy growth while the mean values remain zero. Accordingly, the single maximum of the inclusive rapidity distribution transforms smoothly into the so-called \textit{rapidity plateau}.

The further energy growth makes the maximum points of the summands in \bref{amp_as_sum_again} move away from each other so far that the single maximum of the $T_n \left( \sqrt{s}, X \right)$ observed at lower energies \textit{splits} into a few separate maxima. 
As a result, the eigenvalues $ \lambda_i $ become positive while the first derivatives are still equal to zero, which indicates that there is a local minimum of the function $T_n \left( \sqrt{s}, X \right)$ at the point $\zeta^{(0)}$. It means that as the energy grows, the processes in which the secondary particles have \textit{non-zero} rapidities become more probable than the processes in which all the secondary particles have \textit{zero} rapidities. Hence, the maxima of the inclusive rapidity distribution move away from zero. Moreover, the maximum points of the function $T_n$ are symmetrically distributed about zero due to the symmetry of the considered physical system with respect to the inversion of the collision axis. As a result, the maxima of the inclusive rapidity distribution at high energies are also symmetrically distributed about zero, which is in agreement with the experiment.

\section{Conclusions}

We developed the method of accounting for the interference contributions to an inelastic scattering cross-sections. Using this method, we obtained the models of energy dependence of the total cross-section and rapidity distribution for the proton-proton scattering at energies up to $72$ GeV and multiplicities up to $50$. This models reproduce the experimental data only qualitatively, which can be explained by the fact that the modeling has been performed within the $\phi^3$ (phi-cubed) model -- the simplest dynamical model with the scalar field.

The obtained results indicate that it would be appropriate to use the proposed method with a more complex model, for instance, with the multi-particle-fields model based on QCD \cite{Volkotrub:2015laa}.

In addition, the idea of the proposed method allowed us to analyze the energy dependence of the shape of inclusive rapidity distribution and to propose the physical explanation for this effect.

\bibliography{references}

\end{document}